# Stopping powers of LiF thin films deposited onto self-supporting Al foils for swift protons

Smail Damache[1], Djamel Moussa[2], Saâd Ouichaoui[2,*]

[1] *Division de Physique, CRNA, 02 Bd. Frantz Fanon, B.P. 399 Alger-gare, Algiers, Algeria*
[2] *Université des Sciences et de la Technologie H. Boumediene (USTHB), Faculté de Physique, B.P. 32 El-Alia, 16111 Bab-Ezzouar, Algiers, Algeria*

**Abstract**

The energy losses of ∼ (0.273 − 3.334) MeV protons in LiF thin films deposited by vacuum evaporation onto self-supporting Al foils have been measured using the transmission method. The thicknesses of selected and used LiF/Al target samples were accurately determined via systematic energy loss measurements for alpha particles from a very thin mixed $^{241}$Am/$^{239}$Pu/$^{233}$U radioactive source. The samples were investigated in detail for their stoichiometry and their impurity contents by backscattering Rutherford spectrometry and nuclear reaction analysis. Then, LiF stopping powers have been determined with overall relative uncertainty of less than 2.7% arising mainly from errors in the determination of target sample thicknesses. These $S(E)$ data are reported and discussed in comparison to previous experimental data sets from the literature and to values calculated by the Sigmund-Schinner binary collision stopping theory both for molecular LiF, and for the LiF compound assuming Bragg-Kleeman's additivity rule. Our $S(E)$ data show to be in excellent agreement with the latter theory for molecular LiF over the whole proton energy range explored, which supports the use of modified hydrogenic wave functions for evaluating atomic shell corrections in the case of low-$Z_2$ target materials. In contrast, they exhibit a slightly increasing deviation from theoretical values derived for the LiF compound with assuming stopping force additivity as the proton energy decreases from $E \approx 400$ keV towards lower proton velocities. This deviation in excess relative to experimental data, amounting only up to (at most) ∼ 2.5%, can be ascribed to strong effects of 2s-state valence electrons of Li atoms within the LiF compound. Besides, the comparison to values calculated by the SRIM-2008 computer code indicates that this program satisfactorily accounts for our $S(E)$ data above $E \approx 1.30$ MeV but underestimates them with substantially increasing deviations (up to ~ 11%) towards lower proton velocities where the Bragg-Kleeman additivity rule therefore appears to be inapplicable.

*Keywords:* Vacuum deposition, LiF compound, RBS and NRA analysis, Stopping powers, Bragg-Kleeman's additivity rule.

[*]Corresponding author, Fax : +213 21 24 75 30
E-mail address: souichaoui@gmail.com

# 1. Introduction

The electronic stopping of swift charged particles in matter is a major topic in modern physics research and in various related applications. Recent developments in the evaluation of matter stopping power involving novel experimental methods, theoretical approaches, computer code simulation and tabulations are summarized in Ref. [1]. Presently, the most accurate stopping power data for either elemental or compound targets are required in several fields of fundamental and applied research, such as, e.g., medical and health physics [2-4], materials science and engineering [5, 6] or ion beam analyses [7]. From an experimental point of view, the measurements of the stopping powers of compound and mixture targets are much more difficult to achieve than for elemental targets. Indeed, due to numerous difficulties encountered in preparing and handling the former targets, the precision in their experimental stopping powers is very limited [8] especially when the transmission method is adopted in the measurements. Moreover, the energy loss measurements in composite materials may be severely compromised by ion irradiation effects that may strongly deteriorate and even destroy the fragile structure of these targets. Most often, the latter are described as assemblies of free atoms with conventionally predicting their stopping powers by the Bragg-Kleeman's additivity rule [9] ,i.e., as the simple sums of partial stopping powers of their constituent atoms assumed not to be affected by aggregation effects. However, the atomic overlapping and rearrangement of outer shell electrons usually occur within these target materials. Therefore, aggregation effects including chemical and phase effects may significantly affect the latter leading, sometimes, to important deviations from the additivity rule. This rule is experimentally valid for projectile velocities far exceeding values corresponding to the stopping power function maximum while around and below this region, calculated $S(E)$ values using this assumption show deviations from experiment of ~ 10% up to 13% [10-13]. The dependence of $S(E)$ on the chemical bindings and physical state of the target material is reviewed in reference [14]. Even when related effects are not expected, as for mixtures of non-reactive gases, substantial departures from the strict $S(E)$ additivity rule may occur. Thus, e.g., a deviation from additivity exceeding 50% has been reported in [15] for a He-$H_2$ target mixture bombarded by 8.0 keV deuterons, due to large differences in electron capture and loss cross sections of the two favorably selected gases [16]. Also, conflicting experimental data have been found for solid-gas stopping power differences, and either negative or positive physical state effects were observed [17, 18]. Then, a consistent statement on the Bragg-Kleeman sum rule cannot be made actually. The solid-gas effect has been recently discussed by H. Paul [19]. Besides, recent theoretical investigations [20, 21] based on the binary collision approximation scheme (BCAS) for electronic stopping [22] indicate that significant departures from the $S(E)$ additivity rule may be found for low-$Z_2$ materials, especially light ionic crystals such as lithium fluoride, LiF. Although they may allow further understanding of the stopping powers of compounds, accurate $S(E)$ measurements for swift charged particles crossing these insulator materials are very scarce [8].

In this paper, we report on $S(E)$ measurements for $E \approx (0.222 - 3.312)$ MeV protons passing through LiF thin target films carried out using the transmission method where an overall relative uncertainty lower than 2.7% was achieved. The obtained results are confronted to BCAS [22] and SRIM 2008 code [23] calculations with discussing the validity of Bragg-Kleeman'sadditivity rule. Due to the non feasibility of self-supporting LiF samples with appropriate thicknesses for investigating the above proton energy range, the used LiF films have been deposited onto self-supporting Al foils. As indicated, both these two thin target materials (hereafter called LiF/Al target samples) were carefully prepared by evaporation under high vacuum conditions. Also, a special attention was focused on the preparation and characterization of target samples.

## 2. Experimental procedure and results

### 2.1. Set up

The experimental set up and procedure used in the proton energy loss measurements were described in detail elsewhere [24 - 27]. Therefore, only a brief outline is given below with detailing the preparation and handling of the LiF/Al target samples.

Beams of $H_1^+$, $H_2^+$ and $H_3^+$ ions were delivered by the Algiers 3.75 MV Van de Graaff accelerator through a series of electrostatic lenses and focusing slits with average beam current intensity of ~ 30 nA and typical beam spot diameter of ~ 1.5 mm. They were, first, backscattered off a very thin Au target foil evaporated onto a Si substrate for reducing the beam intensity, then they were used for measuring proton energy losses in transmission through the LiF/Al target samples. Both backscattered ions off the Au target and transmitted ions through the LiF/Al target samples were detected by means of a 500 µm-thick ULTRA ion-implanted Si detector collimated with a ~ 3 mm diameter slit placed at 165° relative to the primary proton beam direction. During experimental runs, the samples were placed onto a movable target holder and could be changed without breaking the high vacuum (pressure of ~ $2 \times 10^{-6}$ mbar) inside the scattering chamber [24]. Prior to performing proton energy loss measurements, the V.d.G. accelerator was calibrated in energy by exploring in details well known resonances from the $^{19}$F (p, αγ) $^{16}$O and $^{27}$Al (p, γ) $^{28}$Si nuclear reactions [28-33]. In this operation, the targets used were prepared by vacuum evaporating pure Al and $CaF_2$ compound onto 0.5 mm-thick Ta backings. For example, in Fig.1 are reported measured yield data from an Al thick target over the resonance at $E_R = 992$ keV in the latter reaction for a collected proton beam charge of 50 µC. Also reported in this figure (solid curve) is the obtained error function best fit to experimental data showing the inflection point at proton energy, $E_p = 991.9$ keV, with energy resolution better than 0.1%.

**2.2. Target samples**

**2.2.1. Preparation**

The LiF material was supplied by "MERCK" manufacturer in form of a powder of very high purity (99.9%) from which LiF/Al samples were prepared by thermal evaporation under good vacuum (pressure < $10^{-5}$ Torr) in three main steps, as described below.

(i) First, we have simultaneously covered polished glass plates with ~ 0.30 µm-thick evaporated potassium bromine layers. The KBr deposits were then used as substrates for subsequently deposited very thin (< 0.2 μm) Al layers. This limit in thickness was selected in order to minimize the proton energy loss and energy loss straggling within these layers by limiting energy loss fractions to at most 5%. Besides, the Al layers had to withstand handling and heat during the LiF deposition process. In addition, advantage was taken of the oxidation of Al in contact with air leading to the formation of a very thin native oxide film upon the surface of Al foils. Indeed, the melting point of this protective Al oxide coating is higher than the LiF boiling point, which confers it a better heat-resistance than metallic Al.

(ii) Then, the double coated glass plates were gradually immersed in distilled water leading to a rapid dissolving of the KBr coating. As a result, the Al films were peeled off and floated on the water surface. The obtained Al films were then picked upon copper mountings, tapped with circular openings of ~ 10 mm in diameter.

(iii) In a last step, the self-supporting Al foils were partly covered with thin LiF layers by partially masking the Al backings during the evaporation process. In this way, the proton energy losses both in the clean Al backings and in the LiF/Al target samples could be simultaneously measured thereby allowing a net determination of proton energy losses in the LiF films (see section 2.2 below). Notice that during this deposition process, no change in the stoichiometry of the evaporated LiF films could occur. Indeed, the dissociation energy of the LiF molecule, $\varepsilon_d = 6.5$ eV [34], corresponds to a temperature, $T \cong 75432$ K, which is very much higher than its boiling point (1717 °C [35]). Then, the LiF molecules surely remain unbroken during the evaporation process. In order to avoid any contamination arising from the Ta boat used, the latter was thermally cleaned by performing a blank evaporation at a temperature considerably higher than that required for evaporating the LiF material but not exceeding its melting point. Several LiF/Al target samples of different thicknesses have been thus achieved and carefully preserved. In order to explore the whole proton energy interval, only two LiF deposit thicknesses of ~ 0.8 µm and ~ 2.0 µm were selected for proton energy loss measurements while the thinnest (~ 0.3 µm-thick) achieved one was used exclusively to probe the elemental composition and impurities of the LiF deposits using ion beams analysis (see below). Notice that all thicknesses were estimated by means of a quartz crystal thickness monitor with relative uncertainty of ~ 20%.

**2.2.2. Thickness determination**

Before carrying out proton energy loss measurements, the selected LiF/Al target samples were investigated for accurately determining their thicknesses through systematic energy loss measurements for alpha particles from a very thin mixed $^{241}$Am/$^{239}$Pu/$^{233}$U radioactive source. These measurements were achieved by using the same experimental arrangement with switching off the proton beam and placing the radioactive source in front of the Au scattering target. This ensured that the same thickness of a scanned LiF/Al target sample was crossed both by the accelerated proton beam and the alpha particles from the radioactive source. Typical experimental alpha particle energy loss distributions recorded for a 2 µm-thick LiF/Al target sample are reported in Fig. 2. In the energy calibration, all adjusted particle peaks in Fig. 2 (a) were used while only the peak of highest energy from the $^{241}$Am radioisotope was considered in the determination of alpha particle energy losses within the LiF layers. Notice that very low rates of carbon and oxygen contaminants of the LiF samples amounting, respectively, to 6.7% and 1.9% (i.e., 3.5% and 1% in atomic concentrations) have been revealed by the Rutherford backscattering spectrometry and nuclear reaction analysis (see below). However, the maximum error induced in the determination of thicknesses of the LiF compound samples (assumed to be pure) amounted to at most 0.4% and could then be considered as negligible. Moreover, these C and O elemental impurities practically do not affect the proton stopping power measurements (see below). Indeed, including these impurities leads to slightly lower stopping power values with differences not exceeding 0.5%. Therefore, the thicknesses of the two investigated LiF deposits were deduced from a combination of measured alpha particle energy loss data and stopping power values generated by the ASTAR program [36] for the LiF compound. Finally, mean areal thickness values of 206 µg/cm$^2$ and 500 µg/cm$^2$ for the two LiF samples were determined with an uncertainty lower than 2.2%. This relative uncertainty in the LiF sample thicknesses mainly results from that affecting the calculated LiF collision electronic stopping power assumed, here, to be of ~ 2%. Indeed, according to the ICRU-49 report [13] supporting the ASTAR program, the uncertainties in tabulated stopping powers are small, ranging between 1% and 2% for elements and between 1% and 4% for compounds and mixtures over the proton high energy region ($E > 0.5$ MeV/amu) where Bethe's stopping theory is very reliable. Making allowance for possible unexpected errors, we have assessed the uncertainty in the target thickness determination to 2.5%. Notice that the full reliability of the above thickness values was also checked by considering the alpha particle peak from the $^{233}$U radioisotope. The thickness values thus deduced were found to be very consistent (to within less than 1%) with those adopted.

### 2.2.3. RBS and NRA analysis

The LiF/Al samples were quantitatively analyzed both for their elemental stoichiometry checking and impurity estimation by RBS scanning under low proton beam intensity (2 up to 6 nA) using the same experimental arrangement as in energy loss measurements with the Au scatter target being replaced by LiF/Al samples. The latter were cooled with a liquid nitrogen trap for reducing carbon deposition. In Fig. 3 are reported RBS spectra (data points shown by solid triangles) for 1300 keV protons backscattered off a 0.3 µm-thick LiF/Al sample with part (a) corresponding to the Al backing alone and part (b) to the Al + LiF coating (with the Al side being oriented towards the incident proton beam). These spectra have been recorded for very low dead time (< 0.04%) and practically the same proton beam fluence (1 up to $2 \times 10^{16}$ cm$^{-2}$). First, we have simulated the Al backing spectrum (see Fig. 3 a) over the ~ (220 – 610) channel range using the SIMNRA program [37]. Making use of the obtained Al baking simulation results, we have adjusted the Al + LiF spectral data over the same channel interval by considering the LiF layer data inputs (thickness, stoichiometry) as the only SIMNRA variable fit parameters. For the simulation of the Al backing RBS spectrum (Fig. 3 a), appropriate stopping powers [38] and non-Rutherford cross sections evaluated by the "Sigma Calc 1.6" software from the IBA Nuclear Data Library (IBANDL) [39] were used. The target sample used in this simulation consisted in a sequence of 4 successive layers, i.e., C-Al$_2$O$_3$-Al(C)-Al$_2$O$_3$ with (C) denoting an amount of homogenously distributed carbon contaminant occurring during the Al vacuum deposition process, the Al$_2$O$_3$ layers consisting in ~ $65 \times 10^{15}$ atoms/cm$^2$ - thick films that arise from air oxidation of the Al surfaces, and C being a ~ $20 \times 10^{15}$ atoms/cm$^2$ - thick layer of amorphous carbon very likely deposited during the Al backing exposition to the proton beam. One can observe (see Fig. 3 a) that the measured RBS spectra and simulated best fit ones (solid curve) are in fair agreement over the entire considered channel interval, thus indicating that the used input stopping power and cross section data are very reliable. As results, a total thickness value of ~ $358 \times 10^{15}$ atoms/cm$^2$ for the Al(C) layer is deduced with a rate of carbon contamination of ~ 3.5%. For the Al + LiF spectrum (Fig. 3 b), the target sample used for simulating the LiF coating consisted in a single layer of LiF(C)(O) mixture with (C) and

(O) denoting small amounts of homogenously distributed carbon and oxygen elemental contaminants occurring during the LiF vacuum deposition and/or the sample handling. Here, cross sections calculated by the "Sigma Calc 1.6" software were used for all but the $^{19}$F and ($^6$Li, $^7$Li) isotopes for which Rutherford cross section and experimental data from references [40, 41], respectively, have been considered. The obtained SIMNRA best fit spectrum to the RBS experimental data over the ~ (220 – 610) channel region is reported in Fig. 3 b. As can be seen, a good agreement is obtained over the whole considered channel interval except for the ~ (480 - 500) range where a noticeable difference is observed, due to the overlapping of peaks from the $^{19}$F(p, p$_1$)$^{19}$F and $^{12}$C(p, p$_0$)$^{12}$C nuclear reactions. Note also that the non-simulated ~ (160 - 210) channel region of experimental data corresponds to the $^7$Li(p,p$_1$)$^7$Li compound inelastic scattering whose cross section data for $E_p = 1300$ keV and the scattering angle $\theta = 165°$ is unknown (as that of the $^{19}$F(p, p$_1$)$^{19}$F scattering), to our knowledge. The initial energies of the p$_1$ protons from these reactions, emitted from near the surface of the LiF coating, precisely correspond to those ($E_{p1} = 953.38$ keV and $E_{p1} = 384.3$ keV, respectively) determined from usual nuclear reaction kinematics for the involved incident proton energy and nuclear reaction Q-values. Finally, the following simulation results have been obtained for the LiF(C)(O) layer: (i) thickness of ~ $3950 \times 10^{15}$ atoms/cm$^2$ and (ii) atomic stoichiometry consisting in elemental 47.75% Li, 47.75% F, 3.5% C and 1% O. One first observes that the same rate in C element impurity was detected both in the Al backing and LiF coating, which was expected since both deposits were elaborated within the same apparatus under similar experimental conditions. Such low rates in light element impurities, typical of the used thermal deposition, have practically un-significant effects on proton energy loss measurements (see next section below), and can then be neglected.

Besides, the detection of low-$Z_2$ element contaminants (such as C, O) has been confirmed by nuclear reaction analysis using a 900 keV deuteron beam. Indeed, Fig. 4 shows the NRA experimental spectrum recorded at laboratory angle $\theta = 150°$ for ion fluence of $\sim 12 \times 10^{16}$ cm$^{-2}$ using the same (0.3 µm-thick) LiF/Al sample. Notice that in this experiment, a ~ 2.5 µm-thick Mylar foil was placed in front of the detector in order to stop scattered deuterons. As can be seen in this figure, the presence of C and O element impurities is clearly evidenced through the detection of three proton groups from the $^{12}$C(d,p$_0$)$^{13}$C, $^{16}$O(d,p$_0$)$^{17}$O and $^{16}$O(d,p$_1$)$^{17}$O nuclear reactions. Also plotted in Fig. 4 is the corresponding SIMNRA simulated spectrum obtained for the same data inputs of the LiF/Al target sample as in the RBS analysis described above (see Fig. 3b) and with taking nuclear reaction cross section data from references [42, 43]. As can be seen, the measured and simulated spectra are in good agreement, attesting for the reliability of the deduced target sample stoichiometry results given above.

**2.3. Stopping power determination and results**

The proton energy losses in the used LiF samples were deduced from measured energy loss distribution spectra such those reported in Fig. 5. Indeed, the difference in mean peak positions for the Al backing considered alone and for the Al + LiF layers yields the mean energy loss, $\Delta E$, within the LiF sample. In a prior step, all measured energy loss distributions were inspected for their possible departures from Gaussian shapes by evaluating the overall dimensionless path-length parameter, $\frac{\Omega_t^2}{T_m^2}$ (see references [24, 44]), associated to each spectrum with $\Omega_t^2$ and $T_m$ denoting, respectively, the experimental energy loss straggling variance and the maximum energy transferred in a single collision for the whole thickness of the LiF/Al target (i.e., the LiF layer + Al backing foil). The $\Omega_t^2$ values were calculated by subtracting in quadrature the measured energy loss distribution variances of the straggled and un-straggled proton beams (i.e., with and without the LiF/Al target in place, respectively). The deduced $\frac{\Omega_t^2}{T_m^2}$ values, ranging between 2.20 and 808.6, substantially exceeded unity for both used LiF/Al target samples. Consequently, neglecting single and quasi-free Coulomb collisions, we have checked that the Gaussian approximation was essentially valid for describing the experimental energy loss distributions.

The deduced mean proton energy loss, $\Delta E$, divided by the average path length, $\Delta x$, crossed by protons within the LiF coating can be considered to within ~ 0.05% [45] as the sample material stopping power, $S_{LiF}$, at the mean proton energy, $E = E_{Al} - \left(\frac{\Delta E}{2}\right)$, for $\Delta E$ amounting to less than 20% of the exit energy, $E_{Al}$, from the Al backing foil. Only in three cases where the energy loss fraction, $\frac{\Delta E}{E_{Al}}$, exceeded 20% (attaining up to ~ 39.83% at the lowest value of the exit energy, $E_{Al} \cong 273$ keV), the $S_{LiF}(E)$ function values were corrected by adding a quadratic correction term derived from an expansion of this function as in reference [45]. However, this second order correction amounts only up to, at most, ~ 0.18%. Besides, the small angle approximation was used to evaluate the proton excess path length within LiF target samples due to multiple scattering [24]. The obtained path length excess ratio versus the exit energy, $E_{Al}$, is reported in Fig. 6 for both LiF samples. As can be seen, the maximum value of this ratio is obtained for the lowest exit proton energy, $E_{Al} \cong 0.274$ MeV, and amounts to ~ 0.07%. This is fully expected since the LiF samples are predominantly constituted of light elements and, then, multiple scattering events are substantially reduced to a negligible level.

The obtained $S_{LiF}(E)$ numerical results are reported in Tab. 1 together with corresponding relative energy loss fractions. Note that these stopping power data are determined assuming pure LiF compound samples. Indeed, the maximum error introduced by this assumption amounts up to at most 0.5% and can be neglected. The overall relative uncertainties in the $S_{LiF}(E)$ data, $\frac{\Delta S_{LiF}}{S_{LiF}}$, were evaluated in the same manner as in our previous work [24]. The major error in the current data arises from the determination of target thickness values (~ 2.5%) while the C and O impurities contribute by ~ 5% to the overall uncertainty.

## 3. Discussion

Our $S_{LiF}(E)$ experimental data are plotted in Fig. 7 where they are compared to previous ones from the literature [46 - 48]. As can be seen, they are in good agreement, within experimental uncertainties, with those measured by Bader et al. [47] and by Biersack et al. [48] for proton energies below ~ 450 keV and ~ 700 keV, respectively. In the remaining higher energy ranges, discrepancies amounting up to ~ 5% and ~ 12% are clearly observed between our data and those reported in references [47, 48]. The $S_{LiF}(E)$ data measured by Draxler et al. [46] exhibit a significant scatter of ~ 5% up to 10% (standard deviation) and lie slightly below our data over the explored common proton energy range. In Fig. 7, our $S_{LiF}(E)$ data are also compared to those compiled in the ICRU-49 report [13]. As can be seen, an excellent agreement between the two sets of data is observed over the high energy region, for $E > 1.0$ MeV, while below this proton energy, our data show to be higher by ~ 4% up to 6%.

In Fig. 8, the measured $S_{LiF}(E)$ data are compared to two sets of theoretical $S_{LiF}(E)$ values calculated by Sharma et al. [20] using the binary collision approximation scheme (BCAS, see [21, 49, 50] and references therein), developed and implemented by P.

Sigmund and A. Schinner in the PASS program [49, 50]: (i) for molecular LiF (i.e., without considering Bragg-Kleeman's additivity rule) and (ii) for the LiF compound based on Bragg-Kleeman's additivity rule. The oscillator frequency spectra database used in the PASS code for the Li, F and LiF materials is obtained as described in references [49, 51] where the oscillator strength spectra are taken from references [52, 53]. For molecular LiF, the calculation is based on a standard procedure where data for inner shell electrons of constituent atoms are used, while the outer shell electrons are collected in one common valence shell. In addition to using orbital binding energies for atoms [54], the authors have adopted ionization potentials for the valence shell from reference [53]. The orbital velocity spectra governing shell corrections are evaluated using modified hydrogen wave functions with an effective charge for each target shell obtained by ionization of that shell [20, 50]. As can be seen, in the case of molecular LiF our $S_{LiF}(E)$ experimental data are fully consistent with the BCAS predicted values over the whole proton energy range explored. This excellent agreement between theory and experimental data not only confirms the validity of the procedure used in the former for obtaining input numerical data for the LiF compound but it also supports the use of modified Hydrogen electronic wave functions for evaluating shell corrections, which are expected to significantly contribute to the departure from the linear stopping force additivity assumption. It must be noted that for evaluating electronic shell corrections, Hydrogen wave functions for low-$Z_2$ target materials up to silver (i.e., $Z_2 \leq 47$) and Hartree-Fock wave functions for high-Z materials ($Z_2 \geq 50$) are used in the ICRU-73 report [50]. It therefore appears that the procedure adopted in the ICRU-73 report for obtaining shell corrections is very consistent for the LiF compound material.

Besides, on can observe in Fig. 8 that for LiF based Bragg-Kleeman's rule, the stopping force additivity hypothesis seems to accurately apply over the proton energy region $E > 400$ keV while an increasing deviation from additivity is noted as the proton energy decreases below this approximate limit, the average deviation amounting only up to ~ 2.5%. This difference is very likely due to valence structure effects of the LiF compound (through electronic structure changes of the 2s state valence electrons in the Li atom) if one assumes that Sharma et al. calculations are less (not) affected by the use of atomic instead of molecular wave functions in computing shell corrections for the LiF molecule. Indeed, the only changes used in the BCAS calculation for this molecule (compared to those based on Bragg's additivity rule) are connected to outermost shell input data (resonance frequency, mean ionization potential) characterizing valence electrons. Moreover, one expects such departure from additivity to occur in the case of low-$Z_2$ compounds like LiF within the projectile low velocity region where excitation channels for inner atomic shell electrons are less involved or even closed while outer shell electrons (whose states substantially differ from those for isolated atomic constituents) should dominate the stopping process.

Also reported in Fig. 8 are $S(E)$-values derived by the SRIM-2008 computer code [23] assuming Bragg-Kleeman's additivity rule. As can be seen, the latter program seems to satisfactorily account for experimental data over the proton energy region above $E \cong 1.3$ MeV while it underestimates the latter for lower proton energies where the stopping force additivity assumption therefore ceases to be valid. A maximum deviation attaining ~ 11% is observed below $E \cong 600$ keV. The observed deviation of the $S(E)$ experimental data from additivity relative to the BCAS predictions is thus much lower than that relative to the SRIM-2008 code calculation, probably due to the fact that low-speed corrections, such as shell corrections for isolated Li and F atoms, are better handled in Sigmund-Schinner binary collision stopping theory where the most adequate up-dated input data (ionization potentials, binding energies, excitation frequencies) are used.

Indeed, in contrast to the BCAS calculation, the shell corrections used in the SRIM code are based on the local density approximation (LDA) making use of Hartree-Fock solid-state charge distributions which probably contribute significantly to the departures from the stopping additivity assumption. Note that for the (low-$Z_2$) LiF compound, about 71% of electrons are located in the outermost atomic shell. Consequently, it appears crucial to correctly take into account all possible complex interaction effects due to these electrons dominating the stopping process, especially at low projectile velocities.

**4. Summary and Conclusion**

In this work, we have accurately measured energy losses for $\sim (0.222 - 3.312)$ MeV protons incident on thin LiF target samples deposited onto self-supporting Al foils and determined corresponding stopping powers with an overall uncertainty of less than 2.7%. First, our $S(E)$ results show to be in excellent agreement with the predictions of Sigmund-Schinner's binary collision stopping theory (BCAS) for molecular LiF over the whole proton energy range explored. This observation thus confirms the adequacy of the standard procedures used in this theory to obtain excitation spectra inputs and modified hydrogen velocity spectra for computing atomic shell corrections for low-$Z_2$ compound targets over the whole covered energy region.

The observed very good consistencies between our experimental data and theoretical or semi-empirical $S(E)$ values calculated by means of several models (ICRU-49, BCAS for LiF compound, SRIM-2008 code) over the high proton energy region (for $E > 1.4$ MeV) attest that the Bragg-Kleeman additivity rule is fully satisfied in this proton velocity region not influenced by valence electron interaction effects, and clearly feature the reliability of our measurements. As noted, a slight difference amounting up to ~ 2.5% is observed between our data and BCAS calculated values with assuming Bragg's additivity below $E \approx 400$ keV, which can be ascribed to valence structure effects of the LiF compound. Indeed, from a theoretical point of view, most pronounced deviations from the simple $S(E)$ additivity for the LiF material have been found [20] for un-screened or weakly screened low velocity projectiles such as antiprotons and protons with respective deviations of ~ 15% and ~ 6% at $E = 100$ keV. Besides, our experimental $S(E)$ data exceed SRIM-2008 computer code predictions assuming additivity by ~ 11 % below $E \approx 600$ keV, then leading one to conclude that for LiF the hypothesis of stopping power additivity is no longer valid for low proton velocities. It must be emphasized that the deviation of our $S(E)$ data from additivity are less important relative to the BCAS approach than to the semi-empirical SRIM code calculation. Finally, it appears that for deriving a more consistent statement on the validity of Bragg-Kleeman's additivity rule and on the role of valence electrons, additional precise stopping power measurements around and below the $S(E)$ function maximum for light ions crossing low-$Z_2$ compound targets, such as light ionic crystals, are highly desirable.

**Figure captions**

**Fig. 1:** Experimental yield curve and corresponding error function best fit for an Al thick target over the resonance at $E_R = 992$ keV in the $^{27}$Al(p, γ)$^{28}$Si nuclear reaction. The collected charge was fixed to ~ 50 μC.

**Fig. 2:** Experimental energy loss distributions (triangles) for alpha particles from a thin mixed $^{233}$U/$^{239}$Pu/$^{241}$Am radioactive source crossing a 500 µg/cm$^2$-thick LiF/Al target sample compared to simulated ones using multi-Gaussian fits (solid curves) to the data: (a) without target sample in place, (b) with placing the Al backing alone and (c) with placing both the LiF and Al layers.

**Fig. 3:** Experimental RBS spectra (triangles) for 1300 keV protons scattered off a 0.3 µm-thick LiF/Al target sample with (a) the Al backing alone and (b) the Al backing + LiF coating. The simulated corresponding best fit spectrum, obtained by the binary collision program SIMNRA [37] over the ~ (220 – 610) channel interval, is also shown (red solid line, see text for details).

**Fig. 4:** Measured NRA spectrum (symbols) recorded at angle θ = 150° for the ~ 0.3 µm-tick LiF/Al target sample exposed to a 900 keV deuteron beam. Also shown is the simulated corresponding spectrum (curve) obtained using the SIMNRA program with the same target data inputs as for Fig. 3.

**Fig. 5:** Typical pulse height spectra for 0.833 MeV backscattered protons: (a) without target sample in place, the P$_0$ peak indicating backscattered protons off the thin Au-layer of the Au-Si target, (b) with placing a 206 µg/cm$^2$-thick LiF/Al target sample, the P$_{Al}$ and P$_{Al+LiF}$ energy loss peaks corresponding, respectively, to the Al backing alone and the Al + LiF layers.

**Fig. 6:** Excess path lengths, $\Delta x/x$, versus the Al-backing transmitted proton energy for both LiF target samples, calculated as in reference [24] using the small angle approximation.

**Fig. 7:** Our measured LiF stopping powers for swift protons (solid circles) versus proton energy compared to previous experimental data from references [46] (open circles), [47] (triangles) and [48] (squares). Our data are also compared to values compiled in the ICRU-49 report [13] (solid curve).

**Fig. 8:** Comparison of the current $S(E)$ experimental data to values calculated by A. Sharma et al. [20] using Sigmund-Schinner's binary collision stopping theory: for molecular LiF (solid curve) and for LiF based Bragg-Kleeman's additivity rule (short-dashed curve). Our data are also compared to $S(E)$ values generated by the SRIM-2008 computer code assuming additivity [23] (dashed-doted curve).

**Table captions**

**Tab. 1:** Measured LiF stopping power, $S_{LiF}$, and corresponding energy loss fraction, $\Delta E/E_{Al}$, versus the mean energy of incident protons.

**Tab. 1**

| $E$ (keV) | $\frac{\Delta E}{E_{Al}}$ (%) | $S_{LiF}(E)$ (MeV cm$^2$/mg) | $E$ (keV) | $\frac{\Delta E}{E_{Al}}$ (%) | $S_{LiF}(E)$ (MeV cm$^2$/mg) |
|---|---|---|---|---|---|
| 222.5 | 39.83 | 0.489 ± 0.013 | 1218.7 | 7.33 | 0.183 ± 0.005 |
| 295.5 | 27.99 | 0.439 ± 0.011 | 1318.6 | 6.40 | 0.172 ± 0.004 |
| 331.8 | 23.76 | 0.412 ± 0.011 | 1418.8 | 5.67 | 0.164 ± 0.004 |
| 386.7 | 19.18 | 0.381 ± 0.010 | 1520.7 | 5.06 | 0.156 ± 0.004 |
| 438.1 | 16.19 | 0.360 ± 0.009 | 1619.8 | 4.55 | 0.149 ± 0.004 |
| 490.7 | 13.70 | 0.338 ± 0.009 | 1719.5 | 4.11 | 0.143 ± 0.004 |
| 542.2 | 11.74 | 0.318 ± 0.008 | 1820.8 | 3.72 | 0.137 ± 0.004 |
| 593.9 | 10.11 | 0.298 ± 0.008 | 1919.4 | 3.38 | 0.131 ± 0.003 |
| 644.8 | 8.78 | 0.280 ± 0.007 | 2118.8 | 2.85 | 0.121 ± 0.003 |
| 697.6 | 7.71 | 0.265 ± 0.007 | 2319.0 | 2.45 | 0.114 ± 0.003 |
| 799.2 | 6.17 | 0.242 ± 0.006 | 2516.5 | 2.12 | 0.107 ± 0.003 |
| 900.7 | 5.09 | 0.224 ± 0.006 | 2716.5 | 1.85 | 0.101 ± 0.003 |
| 949.3 | 4.64 | 0.215 ± 0.006 | 2907.7 | 1.64 | 0.095 ± 0.002 |
| 1048.2 | 3.94 | 0.201 ± 0.005 | 3112.3 | 1.46 | 0.091 ± 0.002 |
| 1147.3 | 3.37 | 0.188 ± 0.005 | 3312.1 | 1.31 | 0.087 ± 0.002 |
| 1118.3 | 8.49 | 0.195 ± 0.005 | | | |

**Fig. 1**

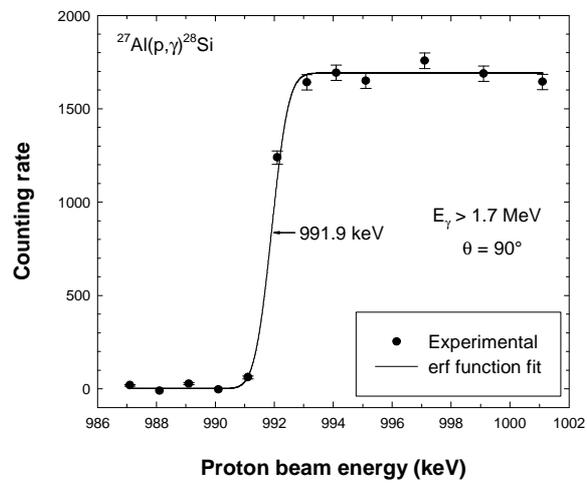

**Fig. 2**

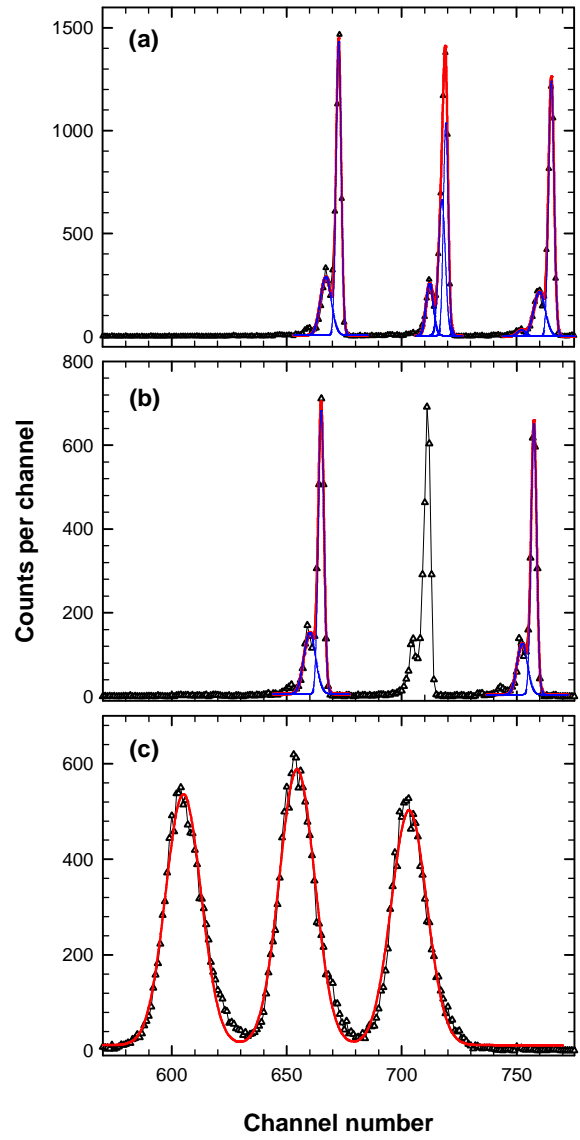

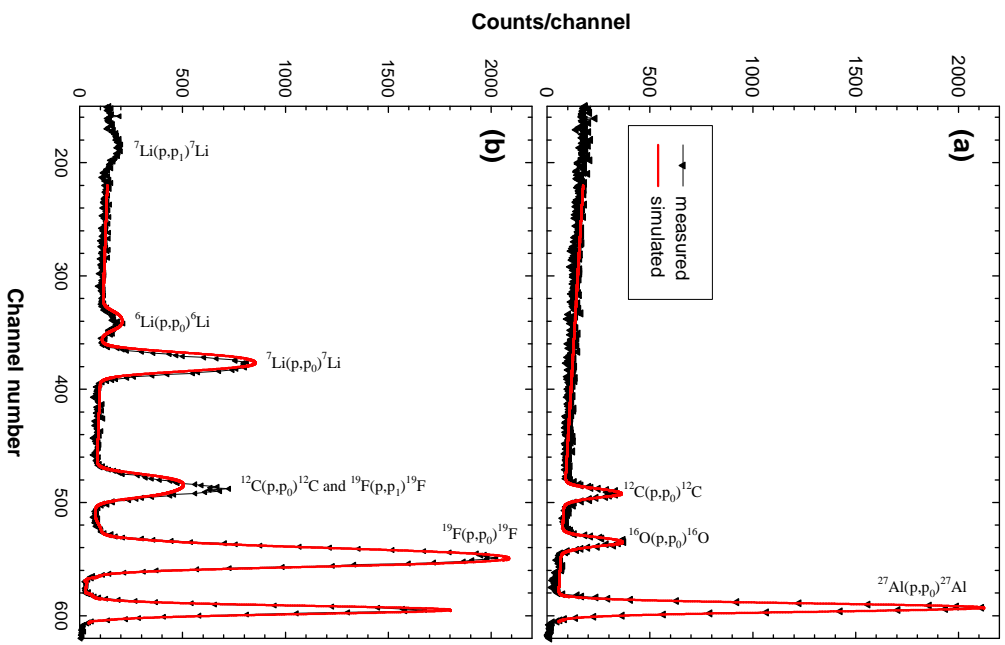

**Fig. 4**

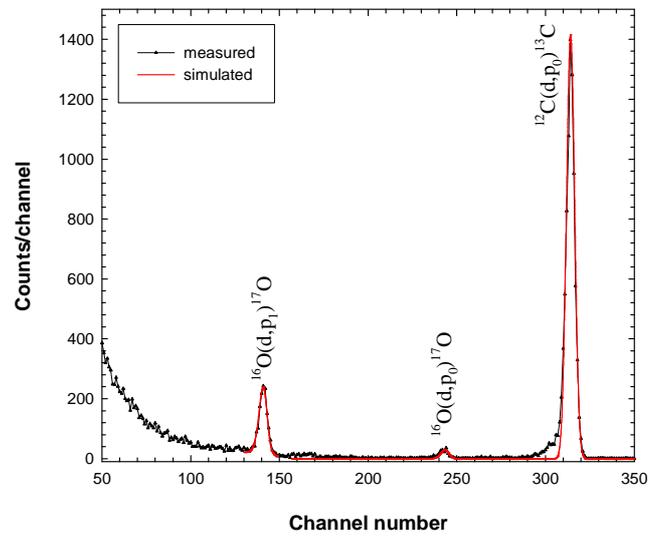

**Fig. 5**

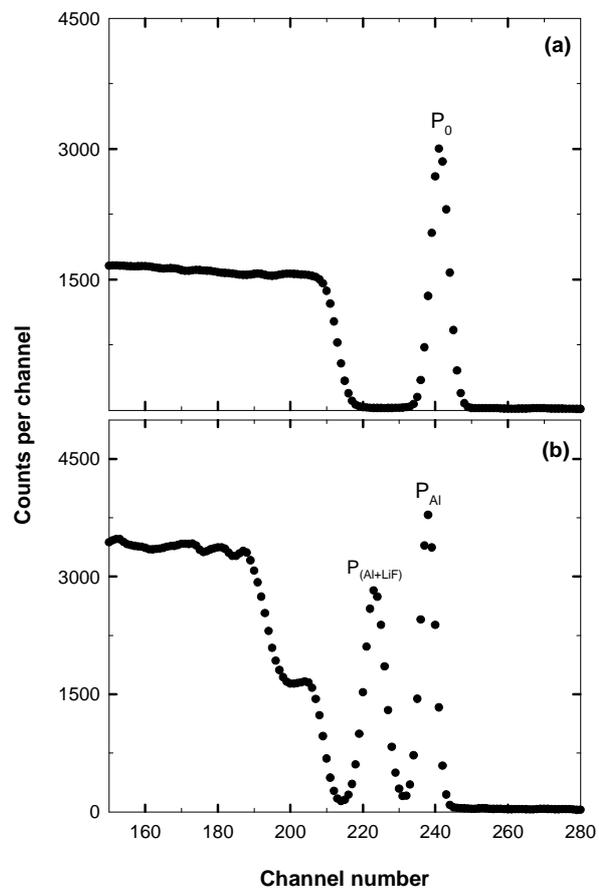

**Fig. 6**

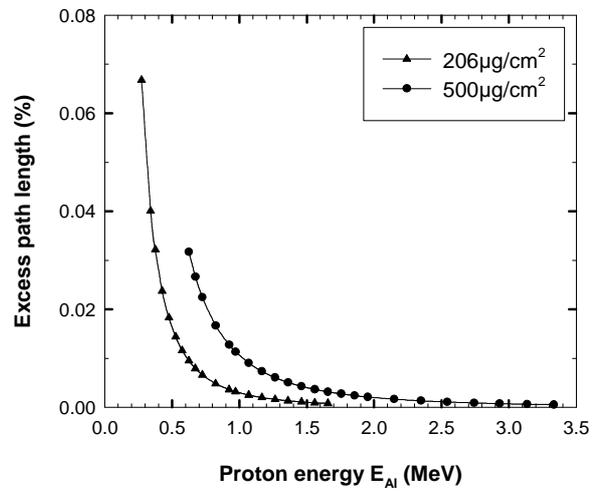

**Fig. 7**

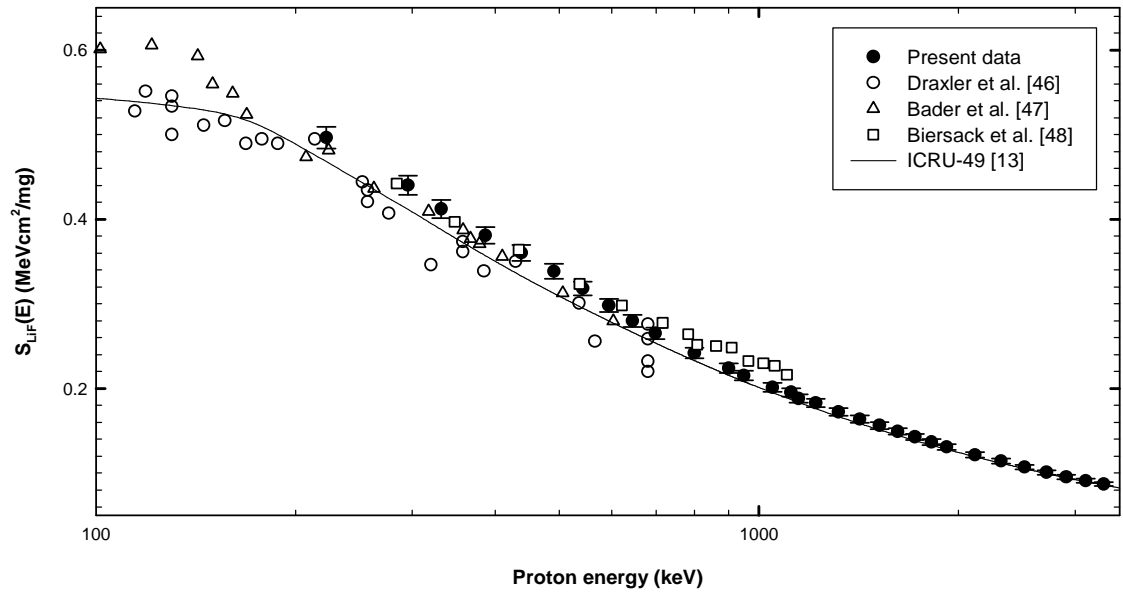

**Fig. 8**

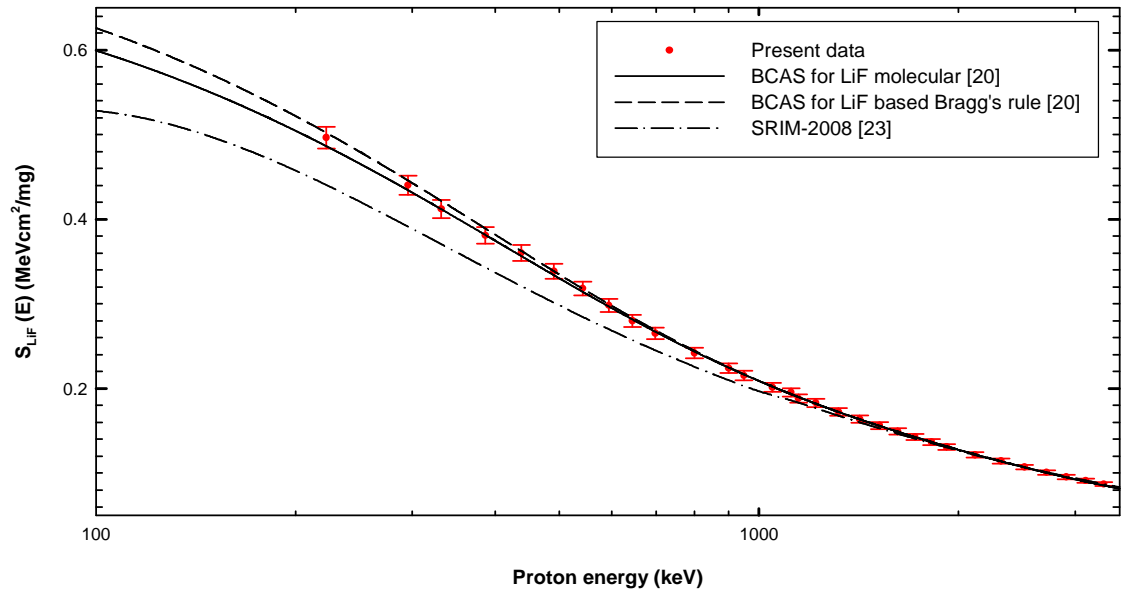